\documentclass[
aip, 
reprint,
superscriptaddress
]
{revtex4-2}

\usepackage[utf8]{inputenc}
\usepackage{siunitx}
\usepackage{graphicx}
\usepackage{amssymb}
\usepackage{booktabs}
\usepackage{color,xcolor}
\usepackage{hyperref}

\begin{document}

\title{Excited-state structure and photoluminescence of the M-center in Si}

\author{Aurora Teien}
\email{aurora.teien@fys.uio.no}
\affiliation{Department of Physics/ Centre for Materials Science and Nanotechnology, University of Oslo, 0316 Oslo, Norway}

\author{David R. Gongora}
\affiliation{Department of Physics/ Centre for Materials Science and Nanotechnology, University of Oslo, 0316 Oslo, Norway}
\affiliation{NNF Quantum Computing Programme, Niels Bohr Institute,
University of Copenhagen, Blegdamsvej 17, 2100 Copenhagen, Denmark}

\author{Arnulf Johannes Snedker-Nielsen}
\affiliation{NNF Quantum Computing Programme, Niels Bohr Institute,
University of Copenhagen, Blegdamsvej 17, 2100 Copenhagen, Denmark}

\author{Viktor Bobal}
\affiliation{Department of Physics/ Centre for Materials Science and Nanotechnology, University of Oslo, 0316 Oslo, Norway}

\author{Augustinas Galeckas}
\affiliation{Department of Physics/ Centre for Materials Science and Nanotechnology, University of Oslo, 0316 Oslo, Norway}

\author{Peter Granum}
\affiliation{NNF Quantum Computing Programme, Niels Bohr Institute,
University of Copenhagen, Blegdamsvej 17, 2100 Copenhagen, Denmark}

\author{Stefano Paesani}
\affiliation{NNF Quantum Computing Programme, Niels Bohr Institute,
University of Copenhagen, Blegdamsvej 17, 2100 Copenhagen, Denmark}

\author{Marianne Etzelm{\"u}ller Bathen}
\affiliation{Department of Physics/ Centre for Materials Science and Nanotechnology, University of Oslo, 0316 Oslo, Norway}

\author{Lasse Vines}
\affiliation{Department of Physics/ Centre for Materials Science and Nanotechnology, University of Oslo, 0316 Oslo, Norway}

\date{\today}

\begin{abstract}
Color centers in silicon offer great possibilities for scalable quantum technologies. The M-center, proposed to originate from a carbon-hydrogen complex, offers telecommunications-band emission and a paramagnetic ground state similar to the T-center. 
Here, we report on photoluminescence lines in the vicinity of the M-center and investigate their properties, including their dependence on temperature, implantation fluence, implantation isotope, and annealing temperature. Three emission lines are observed that are blue-shifted by 1.1, 2.8 and 3.6 meV relative to the 761 meV zero phonon line of the M center, where the 2.8~meV line exhibits negative thermal quenching and is therefore proposed to originate from a second excited state of the M-center. The remaining blue-shifted emission lines, together with two additional red-shifted ($4.7$ and $6.4$ meV) emission lines display normal thermal quenching, and are unaffected by an isotope shift of the implanted carbon atom ($^{12}$C versus $^{13}$C implantation) and implantation fluence. Thus, they likely arise either from other defects with similar emission energies or from a perturbed configuration of the M-center.
\end{abstract}

\maketitle

Color centers in solid-state materials have emerged as promising candidates for quantum technologies (QTs) such as quantum networks and photonic quantum computers \cite{weber_quantum_2010, ruf_quantum_2021}. 
While diamond and silicon carbide (SiC) have been the predominant host materials \cite{ruf_quantum_2021}, silicon (Si) offers significant advantages in terms of scalability and device integration due to its decade-long maturity and the extensive infrastructure of the semiconductor industry. Consequently, Si-based color centers have attracted increasing attention as a platform for large-scale QTs.  
Several color centers in Si have been identified as promising candidates for QT, including the G-center \cite{baron_single_2022, hollenbach_wafer-scale_2022, durand_hopping_2024}, W-center \cite{baron_detection_2022, buckley_optimization_2020}, C-center \cite{wen_optical_2025}, and T-center \cite{bergeron_silicon-integrated_2020, macquarrie_generating_2021, higginbottom_optical_2022}. Other, less extensively studied defects, include the I-center \cite{filippatos_re-examination_2025}, the Al-C center \cite{crosta_bright_2026}, and the recently re-examined M-center \cite{jones_temperature_1973,filippatos_re-examination_2025,filippatos_first-principles_2025}. 
The M-center was first observed optically through photoluminescence (PL) in electron-irradiated p-type Czochralski-grown silicon (Cz Si) in 1973 \cite{jones_temperature_1973}. Under these conditions, the M-center was detected only after annealing at \SI{350}{\celsius} for 20 minutes. Moreover, as annealing temperatures were raised to \(400 \text{ - } 600 \text{ }^\circ\text{C}\), the PL signal diminished, suggesting that the M-center either anneals out or transforms into a dark defect \cite{jones_temperature_1973}. The observed optical transition was attributed to the $^2e \rightarrow {}^2g$ transition between an excited-state doublet and the ground state, which results in a zero-phonon line (ZPL) at 0.760 eV (1631 nm) \cite{jones_temperature_1973}. 

Early studies suggested that the M-center was associated with boron (B) and oxygen (O), as the defect was observed in p-type Cz silicon but not in float-zone Si \cite{jones_temperature_1973}. 
However, subsequent spectroscopic investigations revealed isotope shifts of the M-center zero-phonon line that closely resemble those observed for the T-center, indicating the involvement of carbon (C) in the defect structure \cite{irion_defect_1985}. 
The M-center has been reported to have a triclinic C1 symmetry and introduces an acceptor level at E$_C$-0.37 eV within the silicon bandgap \cite{safonov_m-line_1997}. Based on the experimental observations and recent first-principles calculations, the currently proposed atomic structure consists of three carbon atoms and a single unpaired electron, i.e., a C$_s$C$_i$C$_s$H$_i$ complex \cite{filippatos_first-principles_2025}. Nevertheless, direct experimental verification of this atomic structure remains unconfirmed. 

Theoretical studies further suggest that the M-center shares several key characteristics with the T-center, including a paramagnetic doublet ground state and the presence of an intermediate quartet metastable state that may facilitate spin-selective intersystem crossing \cite{filippatos_first-principles_2025}. These properties could enable optical spin initialization and readout, making the M-center a potentially attractive candidate for spin-photon interfaces and quantum networking applications. The M-center has a reported theoretically calculated radiation lifetime of \SI{1.56}{\micro\second}, which is higher than both the NV-center in diamond (4.4 ns) and the lifetime for T-center (\SI{1.39}{\micro\second}) \cite{filippatos_first-principles_2025}.

A detailed understanding of the origin and electronic structure of the M-center is essential for the development of reliable fabrication and defect-engineering based on this color center. 
In this work, we provide experimental insights on the electronic structure of the M-center by investigating  emission peaks in the vicinity of its main PL line.

We employed a fabrication procedure similar to that commonly used for the T-center, using commercially produced float-zone (Fz) and silicon-on-insulator (SOI) Si wafers\cite{macquarrie_generating_2021, snedker-nielsen_color_2026}. The SOI wafers are bought from Shin–Etsu. 
The samples were first implanted with $^{13}$C ions at an energy of 33~keV. Three implantation fluences were investigated, 5$\times10^{13}$~cm$^{-2}$, 1$\times10^{14}$~cm$^{-2}$, and 5$\times10^{14}$~cm$^{-2}$. Following the carbon implantation, the samples underwent rapid thermal annealing at \SI{1000}{\degreeCelsius} for 20~s in Ar atmosphere. Subsequently, a 140~nm thick sacrificial Al layer was deposited using an electron-beam physical vapor deposition (e-beam PVD) to achieve spatial overlap between the carbon- and hydrogen-implantation profiles. Hydrogen implantation was then performed at 25~keV, after which the Al-layer was removed using KOH. The C:H fluence ratio was maintained at 1:1 following the conditions reported to optimize formation of the T-center \cite{macquarrie_generating_2021}.  Finally, the samples underwent thermal annealing at  \SI{300}{\degreeCelsius} or \SI{575}{\celsius} for 3~min in an N$_2$ atmosphere, in accordance with the two formation regimes reported in Ref. \cite{snedker-nielsen_color_2026}. 
Additionally, a series of samples with identical processing procedure using $^{12}$C ions was produced. 

The full fabrication process, including the implantation series, was done in-house at UiO MiNaLab, except for the final annealing step (\SI{300}{\celsius}, 3~min) that was carried out at the Niels Bohr Institute, University of Copenhagen.

Optical measurements were performed with a photoluminescence (PL) setup consisting of a microscope (10x objective, Mitutoyo) coupled to an imaging spectrograph (Horiba Jobin Yvon, iHR550) equipped with a 300 grooves/mm grating, providing a spectral resolution of $\sim$0.14~nm. The spectra were acquired by a TE-cooled InGaAs detector array (Andor iDus, DU491A). The sample was initially cooled to cryogenic temperatures ($\sim$3.7~K) using a closed-cycle He cryostat and measured at different temperatures using a temperature controller (Oxford Instruments). 
A 400~nm continuous wave (cw) laser (power density of $\sim0.28$~kW/cm$^2$ in the beam spot) was used to optically excite the samples. The laser excitation was directed toward the sample surface at an incident angle of 27$^\circ$ relative to the surface normal. A long-pass filter (LP800) was used to suppress scattered excitation light. A detailed schematic of the PL setup is provided in Ref.~\cite{snedker-nielsen_color_2026}. 

Polarization measurements were performed with a Berek Polarization Compensator in the laser excitation path, which offers tuning of the polarization angle (\SI{360}{\degree}) of the incoming linearly polarized light. 
The samples studied here have (100)-face up, and we excite the sample with an incoming angle of \SI{27}{\degree} relative to the surface normal. Considering the refractive index of Si, the effective angle of the incident beam in the crystal is \SI{6}{\degree}. The incoming laser excitation will thus excite the sample 99.5\% coaxially ($\cos($\SI{6}{\degree}$)$) with a small remaining perpendicular component.


\begin{figure*}[]
    \centering
    \includegraphics[width=\linewidth]{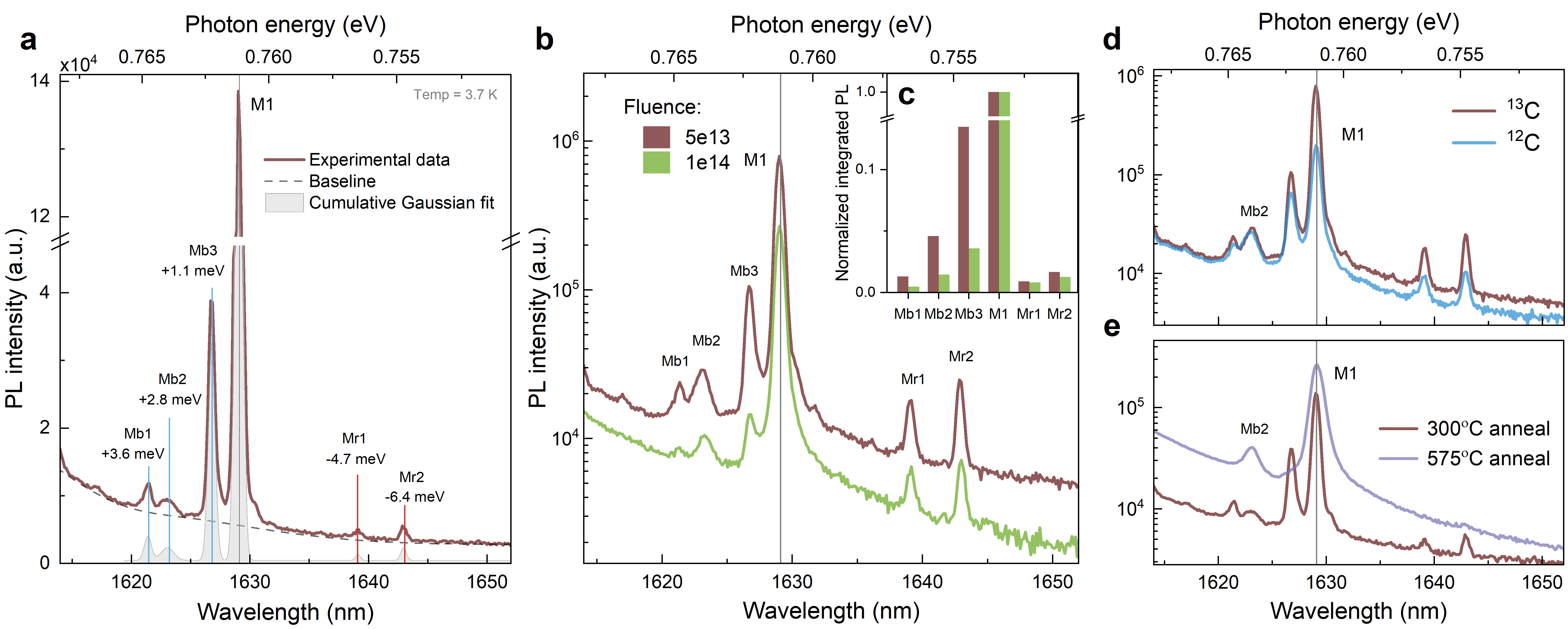}
    \caption{\textbf{PL emission lines in vicinity to the M-center.} (a) PL spectra of the M-center with neighboring blue-shifted satellite peaks Mb1, Mb2, and Mb3, as well as the red-shifted peaks Mr1 and Mr2. The peaks are denoted with the relative energy shift from the M-center ZPL (labeled M1) in meV. The Gaussian fits for all peaks are shown in gray. (b) Fluence-dependence of M-center peaks. The amplitudes of the peaks are fitted by Gaussians and normalized to the amplitude of M1 to compare the relative intensities between the different fluences in (c). (d) PL of samples implanted with different carbon-isotopes. (e) Dependence of M-center emission on the temperature of the final annealing step. Note that the brown curves across all plots are the same, showing PL of a Fz Si sample implanted with $^{13}$C to a fluence of 5$\times10^{13}$~cm$^{-2}$ and a final anneal of \SI{300}{\degreeCelsius}. All measurements are performed at 3.7~K, using a 400~nm excitation wavelength and 10~s exposure time.}
    \label{fig:M_longExp}
\end{figure*}

Figure~\ref{fig:M_longExp}(a) shows a high-resolution PL spectrum of the M-center in a Fz Si sample implanted with $^{13}$C and $^1$H to a fluence of 5$\times10^{13}$~cm$^{-2}$ and with a final annealing temperature of \SI{300}{\celsius}. 
The solid brown line represents the experimental data, which were fit using a combination of the baseline (dashed black line) and a Gaussian function for each peak (shaded gray areas). The fit was used to determine the central wavelength (cwl), area, and amplitude of each peak. Similar peaks were also observed in the SOI sample annealed under the same conditions.
The main M-center ZPL, labelled M1, is located at 1629.0~nm ($\sim761.2$~meV). Interestingly, long-exposure PL measurements reveal a plethora of features in the vicinity of the M-center (Fig. ~\ref{fig:M_longExp}(a)). In addition to the main M-center ZPL, two red-shifted emission peaks are observed at energy offsets of -4.7~meV and -6.4~meV, while three blue-shifted peaks appear at +1.1~meV, +2.8~meV, and +3.6~meV relative to the M-center ZPL. The red-shifted peaks are labelled Mr1 and Mr2, whereas the blue-shifted peaks are labelled Mb1, Mb2, and Mb3 (Fig.~\ref{fig:M_longExp}(a)). Indeed, the presence of Mb2, Mr1, and Mr2 has been reported in previous work \cite{safonov_m-line_1997}. 
A similar manifold of related PL lines has been reported for the C-center \cite{wen_optical_2025}, where the additional peaks were attributed to distinct excited states arising from the excitonic nature of the defect.

In Fig.~\ref{fig:M_longExp}(b), PL spectra of the M-center manifold are presented for samples processed under identical conditions, except for the implantation fluences: 5$\times10^{13}$~cm$^{-2}$ (brown curve) and 1$\times10^{14}$~cm$^{-2}$ (green curve). Samples implanted with an even higher fluence of \SI{5e14}{\per\centi\meter\squared} were also investigated but exhibited no detectable M-center emission.
This may indicate that a higher implantation fluence is not favorable for M-center formation or that it promotes the formation of competing defects and/or non-radiative recombination centers that inhibit PL of the M-center. 
Notably, for both fluences shown in Fig.~\ref{fig:M_longExp}(b), all the same peaks are present. Mapping the amplitudes of the peaks using Gaussian fits and normalizing it to M1, we obtain the inset barplot in Fig.~\ref{fig:M_longExp}(c). Visibly, the blue-shifted peaks are more prominent in the lower fluence sample (5$\times10^{13}$~cm$^{-2}$). Indeed, the peaks Mb1, Mb2 and Mb3 are respectively $\sim$2.8, $\sim$3.2, and $\sim$3.8 times more prominent for the lower fluence. In contrast, the red-shifted peaks show no significant change depending on fluence; the ratio is $\sim$1.1 for Mr1 and $\sim$1.3 for Mr2. 
As a result of this fluence-sensitivity, we expect the red-shifted peaks to be related. 

In Fig.~\ref{fig:M_longExp}(d), results from the $^{13}$C isotope implantation are compared to those of the  $^{12}$C implant, where the same neighboring peaks are observed at approximately the same wavelength. Note that the isotope shift of the ZPL reported in \cite{irion_defect_1985} is only a few tenths of $\mu$eV and not visible in the present measurements. Fig.~\ref{fig:M_longExp}(e) displays the PL spectra of samples with different annealing temperatures, i.e., \SI{300}{\degreeCelsius} or \SI{575}{\degreeCelsius}. Interestingly, only M1 and Mb2 appear in the sample annealed at the higher temperature.

\begin{figure*}[t]
    \centering
     \includegraphics[width=\linewidth]{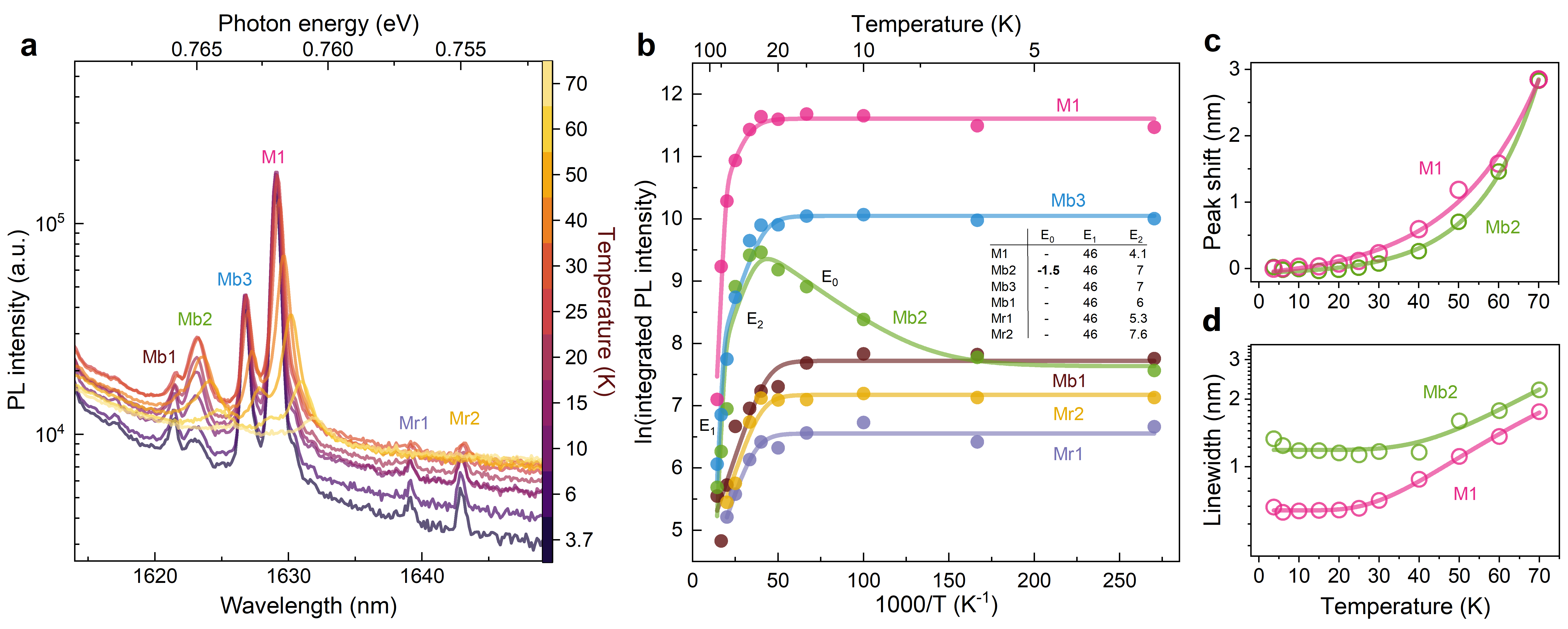}
    \caption{\textbf{Temperature dependence of M-center emission.} (a) PL temperature series of the M-center ZPL (M1) and neighboring peaks (Mb1, Mb2, Mb3, Mr1, and Mr2) in Fz Si implanted with $^{13}$C and $^{1}$H. (b) Arrhenius plot showing the dependence on PL collection temperature of the M1 peak and its neighbors. The solid curves in (b) are numerical fits of the experimental data (dots). 
    A 1.5~meV activation energy is identified for Mb2 from the slope of the linear fit, evidencing negative thermal quenching. The remaining activation energies are included in the table inset. The shift and broadening of the Gaussian peak fits for M1 (pink) and Mb2 (green) are shown in (c) and (d), respectively. The solid lines in (c) are exponential fits, while the solid lines in (d) are fits from a simple thermally induced transition model. The temperature series measurements were conducted with 400~nm excitation at an incident angle of \SI{27}{\degree} relative to the surface normal. 
  }
    \label{fig:tempseries}
\end{figure*}

For further insights and identification of the satellite peaks, we perform a temperature-dependent PL sweep. 
Fig.~\ref{fig:tempseries}(a) shows a temperature series from 3.7~K to 70~K of the M-center ZPL, denoted M1 in pink, and the neighboring peaks, Mb1 (brown), Mb2 (green), Mb3 (blue), Mr1 (violet), and Mr2 (yellow). The variation in integrated PL intensity of the peaks as a function of temperature is shown in the Arrhenius plot in Fig.~\ref{fig:tempseries}(b). Here, the solid lines represent numerical fits to the data (data points), from which the activation energies can be extracted from the slopes of the linear regions. 

Figure~\ref{fig:tempseries}(a) and (b) reveal that all peaks undergo thermal quenching at temperatures above 30~K. However, one of the adjacent blue-shifted neighbouring peaks (Mb2) experiences an initial negative thermal quenching at low temperatures ($<$30~K) before the normal quenching takes over at higher temperatures. 

The shift in cwl and linewidth for M1 (pink) and Mb2 (green) as a function of PL collection temperature are shown in Fig.~\ref{fig:tempseries}(c) and \ref{fig:tempseries}(d), respectively. The two PL signatures follow the same exponential temperature-dependent shift in cwl (Fig.~\ref{fig:tempseries}c) consistent with the temperature dependence of the bandgap of Si \cite{yoo_temperature_2015}. 
Further, Fig.~\ref{fig:tempseries}(d) displays the temperature dependence of the PL linewidth of M1 (pink) and Mb2 (green). The experimental data are fitted with a thermally induced transition model. 
From the fits, we extract thermal activation energies of 11.9(6)~meV and 18(4)~meV for M1 and Mb2, respectively.

The implantation fluence dependence initially suggests that some of the satellite peaks could be related to the M-center. However, only the Mb2 is detectable in the \SI{575}{\degreeCelsius} annealed sample. 
Interestingly, the M-center was not found in the $^{13}$C implanted Fz sample annealed at \SI{575}{\celsius}, only in the $^{12}$C implanted SOI sample annealed at \SI{575}{\celsius} (violet curve in Fig.~\ref{fig:M_longExp}(e)).

Note also that the M1 peak width is substantially broader in the \SI{575}{\degreeCelsius} sample. 
This broadening may indicate a modification of the M-center configuration or its local environment induced by the higher annealing temperature. Similar effects have been observed for the G-center, where atomic reconfiguration and local perturbations, particularly strain in SOI, were shown to influence the emission fine structure strongly \cite{durand_hopping_2024}. 
Nonetheless, further experiments are required to fully disentangle the origin of the emissions.

The peaks labelled Mr and Mb herein could also originate from other M-like defects in Si. 
Red-shifted satellite peaks, such as Mr1 and Mr2, are often attributed to the phonon sideband or local vibrational modes. Emission from local vibrational modes is expected to be mass dependent, i.e., exhibit a shift in wavelength if a different isotope is used. Indeed, this has been observed for the T-center \cite{bergeron_silicon-integrated_2020, kazemi_giant_2026}. The PL spectra in Fig.~\ref{fig:M_longExp}(d) reveal no shift between the $^{13}$C and $^{12}$C implantations, supporting an assignment of Mr1 and Mr2 arising from perturbations or different configurations. Similar arguments could be stated for Mb1 and Mb3.

Expected additional M-center features, such as phononic transverse acoustic (TA) and transverse optical (TO) lines as well as local vibrational modes (LVM), are not investigated in this work due to the rapidly decreasing quantum efficiency of the detector used at wavelengths above 1650~nm. 
However, detailed investigations of the M-center LVMs have previously been reported in Ref. \cite{safonov_m-line_1997}. Here, two red-shifted peaks close to the M-center (assumed to correspond to Mr1 and Mr2 of this work) are reported to have their own local mode satelites around 680~meV and 690~meV, which further supports the hypothesis of their M-center independence.

Mb2 exhibits a negative thermal quenching behaviour visible as an initial increase in integrated PL intensity for lower to intermediate temperatures in Fig.~\ref{fig:tempseries}(b). Further, the thermal broadening of the M1 line in Fig.~\ref{fig:tempseries}(d) picks up earlier than for Mb2, possibly due to the thermally activated transitions between them. This is in accordance with a model where Mb2 stems from a higher-lying excited state of the M-center, similar to what has been found previously for, e.g., silicon vacancy-related defects in 4H-SiC \cite{bathen_resolving_2021, rubin_seeing_2026}.  The position of this excited state at 2.8 meV higher than M1 is consistent with previous findings by Safonov et al. \cite{safonov_m-line_1997}, where Mb2 and M1 are interpreted as two Kramers doublets whose degeneracy is lifted by the defect core potential. 
The thermal activation energy for the excited state is found to be 1.5~meV in this work, which is close to the energy spacing between the TX0 and TX1 bound-exciton energy levels ($\sim1.76$~meV) of the T-center.

\begin{figure}[h]
    \centering
    \includegraphics[width=0.97\linewidth]{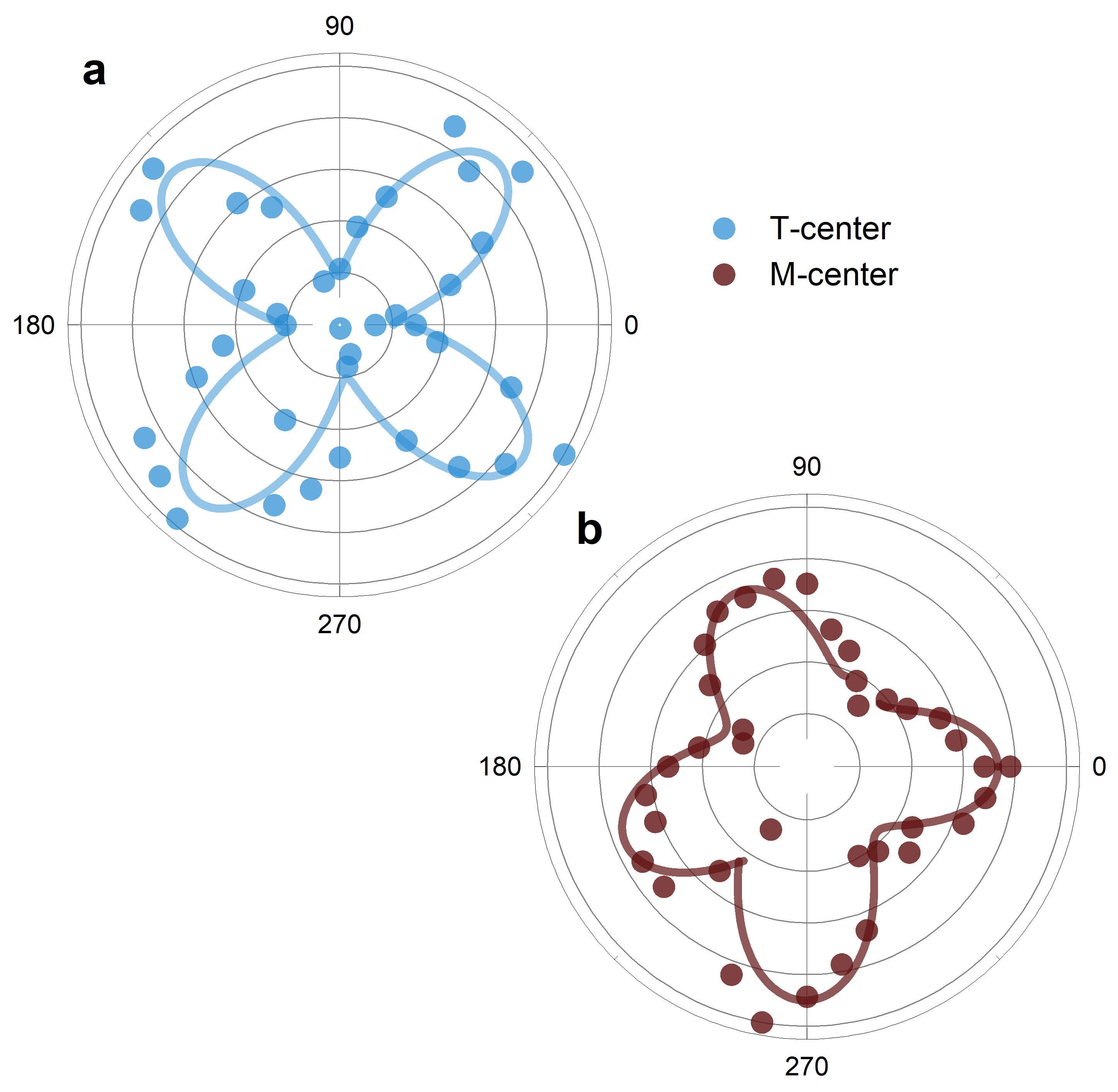}
    \caption{ \textbf{Polarization dependence of ensemble T- and M-centers.} Incoming polarization angle dependence of observed PL intensity for the (a) T-center (blue) and (b) M-center (brown). More specifically, we track the TX0 and M1 peaks. The PL is performed with a 400~nm excitation source at 3.7~K. The data is fit using Malus' law.  }
    \label{fig:polarization1}
\end{figure}

Given that the predicted structure of the M-center, C$_s$C$_i$C$_s$H$_i$ \cite{filippatos_first-principles_2025}, holds true, then the M-center has a low symmetry (triclinic, $C_1$) and would therefore be expected to exhibit polarization-dependent absorption and emission. 
The related T-center has shown $C_{1h}$ symmetry, which offers slightly higher symmetry than the proposed M-center structure. 
However, an ensemble of the T-center can have 12 different inversion symmetry orientations, considering the crystal axis \cite{xiong_computationally_2024, clear_optical_2024}. The theoretically predicted structure of the M-center is supposed to be located in a similar way in the Si host crystal as the T-center, but with additional degrees of freedom (72 possible configurations). 
As an initial experiment, the PL intensity as a function of incoming laser polarization angle is shown for T- and M-center ensembles in Fig.~\ref{fig:polarization1}. Visibly, the T-center exhibits a four-fold symmetry. Indeed, for an ensemble of single-emitters in Si(100), the expected equivalent contribution from two orthogonal (001) and (010) planes would imply 4-fold symmetry of PL polarization dependence in a polar plot. 
Indeed, the M-center generally exhibits 4-fold rotation anisotropy. Note that the data were fitted using Malus' law, $I(\theta)= A + B \cos2(\theta-\alpha)$, to correctly map the polarization response of the defect. 
The results indicate a polarization dependence of the absorption, which is important for a conclusive determination of the defect identification. 

The measurements reveal distinct polarization-dependent PL responses of the T- and M-center under 400~nm excitation. Because this excitation is non-resonant with the ZPL transitions, the observed modulation does not directly stablish the M1 or TX0 absorption rules or transition dipole orientations. The observed modulation may involve anisotropic higher-energy defect excitation and/or carrier relaxation and capture process. Further work is needed to identify both the structure of the M-center and the directionality of its absorption dipole.


Experimental formation methods developed for the T-center in Si were employed to produce the M-center, a telecom-emitting defect with a ZPL in the L-band. The resulting defects were characterised using photoluminescence spectroscopy, revealing a series of satellite peaks in the vicinity of the M-centre ZPL. 
Temperature-dependent measurements reveal a thermal activation energy of 1.5~meV for the higher-lying excited state, whose emission intensity reaches a maximum at 30~K.
These results provide insights into the electronic structure and optical emission of the M-center in Si, as well as the spectral features in its immediate vicinity, providing important information for evaluating its potential for integration into scalable quantum technologies.


\acknowledgements 
Financial support was kindly provided by Akademiaavtalen between Equinor and the University of Oslo through the research project QSenS, and by the Research Council of Norway through the Centre for Defects in Semiconductors for Quantum Sensing (Project No.~354831) and the Norwegian Micro- and Nano-Fabrication Facility, NorFab, Project No.~349807. 

Additional financial support was provided by the Novo Nordisk Foundation, Grant number NNF22SA0081175, NNF Quantum Computing Programme. S.P. acknowledges funding from VILLUM FONDEN (Grant No. VIL60743 and VIL78724) and the European Research Council (ERC StG ASPEQT, No. 101221875).

\section*{Author declarations}
The authors have no conflicts to disclose.

\section*{Data availability}
The data that support the findings of this study are available from the corresponding authors upon reasonable request.

\section*{References}

\bibliographystyle{naturemag}
\bibliography{sample}

\clearpage


\end{document}